\documentclass[
reprint, 
bibnotes,
amsmath,amssymb,
aps,
prx,
reprint
]{revtex4-2}

\usepackage{graphicx}
\usepackage{dcolumn}
\usepackage{bm}
\usepackage{hyperref}
\usepackage{physics}
\usepackage{xcolor}
\usepackage{dsfont}

\begin{document}

\preprint{APS/123-QED}

\title{\textbf{Metrology of quantum imaging schemes} 
}%

\author{Emma Brambila}
\author{Giacomo Sorelli}%
 \email{Contact author: giacomo.sorelli@iosb.fraunhofer.de}
\affiliation{%
Fraunhofer IOSB, Ettlingen, Fraunhofer Institute of Optronics, System Technologies and Image Exploitation\\
Ettlingen, Germany
}%

\date{\today}

\begin{abstract}
We compare the performance of quantum imaging schemes based on spatially correlated photon pairs by formulating them as quantum multiparameter estimation problems, in which the object is characterized by transmission coefficients associated with different spatial modes. Our work focuses on standard quantum imaging techniques such as ghost imaging, two-photon imaging, and imaging with undetected photons. Specifically, we compute the quantum Fisher information matrices and show that they are saturated by Fisher information matrices corresponding to measurements in the object-mode basis, which generalize the common detection schemes employed in each imaging configuration.
We find that ghost imaging and two-photon imaging generally provide higher precision for transmission estimation than imaging with undetected photons, but the latter is the only one that naturally does not couple transmission estimation across different spatial modes.
These results identify which imaging protocols are best suited for specific tasks and provide practical guidelines for the design and optimization of quantum sensing technologies based on spatial correlations.
\end{abstract}


\maketitle

\section{Introduction} 
Quantum imaging schemes exploit correlations between photons to probe objects in regimes in which conventional imaging is limited \cite{Defienne2024}. These include low‑light applications, where the photon flux must be reduced to avoid damaging sensitive samples \cite{Gemmell2023}; imaging through intrinsically high‑loss environments \cite{Kim2021}, as well as wavelength‑mismatched scenarios, where the object must be illuminated within a specific spectral range for which off‑the‑shelf detectors either perform poorly or are impractically bulky \cite{Lemos2014,Inna2022}. Beyond these technological advantages, it has also been shown that, in specific cases, correlated photon pairs can enable higher spatial resolution \cite{Defienne2022,Milena2001} and novel applications such as image hiding \cite{Chloe2024} and quantum adaptive optics techniques \cite{Patrick2024}.

Despite the variety of quantum imaging protocols that have been proposed and experimentally demonstrated, their performance is often assessed using heuristic metrics—such as visibility \cite{Kolobov2017}, case-specific resolution \cite{Moreau2018,Jorge2022}, or signal-to-noise ratio \cite{Kim2021}—under diverse modeling assumptions. As a result, it is difficult to compare different schemes performance.

In parallel, quantum metrology provides a general framework for quantifying the ultimate precision with which parameters can be estimated from specific quantum states of light \cite{Paris2009,Giovannetti2011}. These tools have already been applied to specific imaging problems, such as estimating light-source positions \cite{Treps:2003,Delaubert:2006,Boeschoten:26} and separations \cite{Tsang2017,Tan2023,Rouviere:24}.
In this work, we bring together quantum imaging and quantum metrology by treating imaging methods as parameter-estimation problems for object transmission characteristics. 
Specifically, we analyze the most common quantum imaging techniques: ghost imaging \cite{Strekalov1995,Milena2001,Kim2021}, two-photon imaging \cite{Defienne2022,Chloe2024,Patrick2024}, and imaging with undetected photons \cite{Lemos2014,Inna2022,Gemmell2023}, within a unified framework based on the fundamental precision bounds from quantum metrology \cite{Liu2020}. This approach allows a systematic comparison of their strengths and limitations for object transmission characterization.

To introduce this analysis, Sec.~\ref{sec:intro} develops the theoretical framework. We first describe the structure of spatially correlated photon-pair generated by spontaneous parametric down-conversion (Sec.~\ref{sec:intro_SPDC}), then show how the imaging problem can be mapped onto a multiparameter estimation problem for the transmission coefficients characterizing the object (Sec.~\ref{sec:intro_obj}) and finally recall the quantum metrology tools that are needed throughout the remainder of the paper (Sec.~\ref{sec:intro_Metro}). 
We present the results for each imaging technique in Sec.~\ref{sec:All_img} and compare them in Sec.~\ref{sec:Comparison}.

We derive the ultimate precision limits for reconstructing an object's transmission profile and identify the measurements that attain them. Our analysis shows that the attainable precision for characterizing an object depends on how the probe state's correlations are distributed among the object's transmission modes, providing a unified framework to compare different quantum imaging protocols. In particular, this framework reveals clear performance differences between the considered techniques in different transmission regimes and indicates which schemes are best suited to different tasks, thereby providing practical guidelines for selecting and optimizing quantum imaging strategies for specific applications.
 
\section{From Biphoton Imaging to Quantum Estimation}
\label{sec:intro}
\subsection{\label{sec:intro_SPDC} Spontaneous parametric down-conversion}
Most quantum imaging schemes rely on Spontaneous Parametric Down-Conversion (SPDC): a non-linear process in which a pump photon spontaneously converts  into a pair of lower-energy correlated photons, conventionally labeled as signal ($s$) and idler ($i$).
In the paraxial, semi-monochromatic regime, this process produces the spatially multi-mode photon state \cite{Walborn2010}

\begin{equation}
|\psi\rangle_\textrm{SPDC}= \iint_{-\infty}^{\infty}d\vec{q}_sd\vec{q}_iC(\vec{q}_s,\vec{q}_i)|\vec{q}_s\rangle |\vec{q}_i\rangle.
\end{equation}
Here $\ket{\vec{q}_s} = \hat{a}^\dagger_{\vec{q}_s}\ket{0}$ ($\ket{\vec{q}_i} = \hat{a}^\dagger_{\vec{q}_i}\ket{0}$) denotes the quantum state of a signal (idler) photon with transverse wave vector $\vec{q}_s$ ($\vec{q}_i$), and the {\it joint spatial amplitude}  
$C(\vec{q}_s,\vec{q}_i)$ defines the joint probability $P(\vec{q}_s,\vec{q}_i)=|C(\vec{q}_s,\vec{q}_i)|^2$ 
of detecting signal and idler photons with transverse wave vectors $\vec{q}_s$ and $\vec{q}_i$.

To characterize the spatial entanglement of the SPDC state, we introduce a Schmidt decomposition of the joint spatial amplitude \cite{Miatto2012}
\begin{equation}
  C(\vec{q}_s,\vec{q}_i)
  =\sum_{n=0}^\infty \sqrt{\lambda_n}\,
  u_n(\vec{q}_s)\, v_n(\vec{q}_i),
\end{equation}
with $\{u_n\}$ and $\{v_n\}$ being orthonormal sets of signal and idler modes, respectively, and the Schmidt coefficients $\lambda_n \ge 0$ such that $\sum_{n=0}^{\infty} \lambda_n = 1$. 
Therefore, the bi-photon SPDC state in the Schmidt basis can then be written as 
\begin{equation}\label{eq:SPDC_Schmidt}
  |\psi\rangle_\textrm{SPDC}
= \sum_{n=0}^{\infty} \sqrt{\lambda_n}\;
  |u_n\rangle_s\,|v_n\rangle_i,
\end{equation} where 
$|u_n\rangle_s=\int d\vec{q}_s\, u_n(\vec{q_s})\,\ket{\vec{q}_s},$ 
$|v_n\rangle_i=\int d\vec{q}_i\, v_n(\vec{q_i})\,\ket{\vec{q}_i}.$

Although the Schmidt decomposition is technically infinite dimensional, in practice only a finite number of Schmidt coefficients give a significant contribution. This effective dimensionality of the state is quantified by the Schmidt number
$K=1/\sum_{n=0}^{\infty} \lambda_n^2$ \cite{Law2004}. 
Accordingly, the sum in Eq.~\eqref{eq:SPDC_Schmidt} can often be truncated to roughly $K$ terms without appreciable loss of accuracy.
From a quantum imaging perspective, the Schmidt number determines the effective spatial dimensionality of the probe state, implying that object features encoded in spatial modes of order much larger than $K$ cannot be resolved.

\subsection{\label{sec:intro_obj}Encoding of object information}

Any linear optical object admits a singular-value decomposition in terms of input and output singular mode bases, $\{\ket{f_k}\}$ and $\{\ket{g_k}\}$ \cite{Popoff2010,GolubVanLoan2013}. In this representation, the interaction with a single photon prepared in mode $\ket{f_k}$ is described by \cite{Lemos2014}
\begin{equation}\label{eq:object}
\ket{f_k} \;\rightarrow\; \tau_k \ket{g_k} + \sqrt{1-\tau_k^2}\,\ket{e_k},
\end{equation}
where $\tau_k \in [0,1]$ are the singular values associated with transmission channels describing the structure of the object, and $\{\ket{e_k}\}$ are orthogonal environmental modes representing the lost component.

\begin{figure}[t]
\centering
\includegraphics[width=0.45\textwidth]{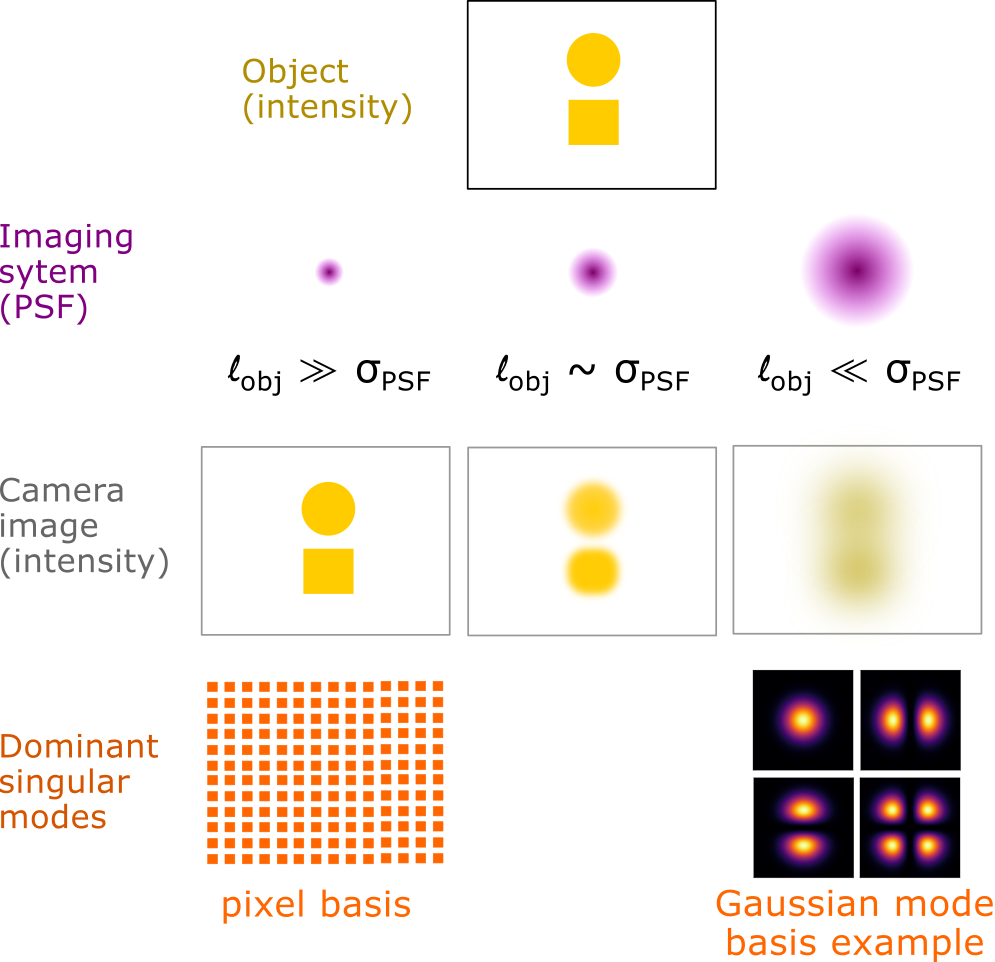}
\caption{Effect of the point-spread function (PSF) widths $\sigma_{\rm PSF}$, corresponding to different singular-mode decompositions, from a fixed object of size $\ell_{\rm obj}$.}
\label{fig:SV}
\end{figure}

\begin{figure*}[ht!]
\centering
\includegraphics[width=0.67\textwidth]{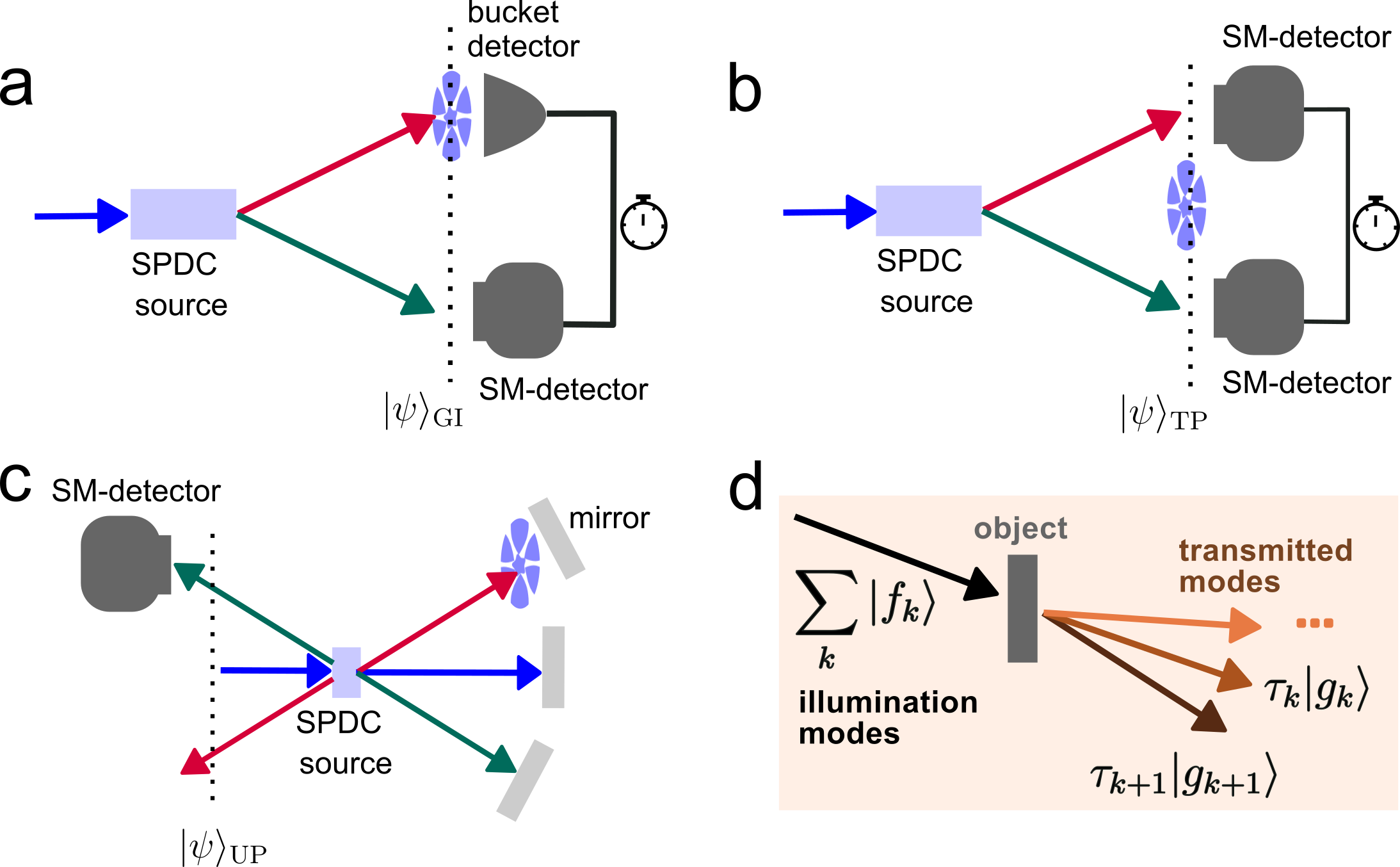}
\caption{Simplified experimental schemes for different imaging techniques based on SPDC spatial correlations: a) ghost imaging, b) camera-based coincidence detection, and c) imaging with undetected photons. We highlight that the measurement for each technique can differ depending on which singular modes are resolved by the detector (SM-resolving detector). The dashed line shows where the state is written down. In d) we illustrate how our metrology analysis is based on a singular-mode decomposition, where the illumination of an object is a superposition in the object basis and the output is described by the transmission coefficients $\tau_k$.}
\label{fig:QI_all}
\end{figure*}

In general, a complete characterization of an arbitrary optical object requires determining both its singular values and its associated singular modes. There are, however, some imaging regimes in which the singular-mode structure is approximately known from the imaging system through its point-spread function (PSF), which defines the characteristic resolution scale (see Fig.~\ref{fig:SV}). When the object features are much larger than this scale, the imaging operator is approximately diagonal in the position basis, and the singular modes reduce to localized pixel modes. Conversely, in the sub-diffraction regime, where only a few closely spaced point-like features are relevant, the singular modes are well approximated by PSF-adapted spatial modes determined by the optical system. In this work, we focus on these regimes and assume the corresponding singular-mode basis to be known, restricting the estimation problem to the singular values ${\tau_k}$ (see Fig.~\ref{fig:QI_all}\textbf{d}). Imaging scenarios in which the singular modes themselves must also be inferred are beyond the scope of the present treatment.
For simplicity, we will refer to the object's singular modes simply as object modes throughout the remainder of the paper.

In the case of biphoton states, different scenarios of object interaction and measurement schemes have been implemented, see Fig.~\ref{fig:QI_all}\textbf{a-c}. Even though all these methods rely on spatial correlations, they differ significantly in both their detection strategies and how the object information is encoded \cite{Defienne2024}. Ghost imaging (GI) exploits temporal coincidence detection to extract spatial information from the photon that did not interact with the object. The two-photon imaging approach (TP) instead relies on the shared information carried by photon pairs that simultaneously illuminate the object, with the image reconstructed from coincidence measurements. Finally, imaging with undetected photons (UP) detects only the photon that does not interact with the object, while this information is inferred through induced interference.

Accordingly, we consider measurements that resolve the relevant object modes. In the resolved regime, this corresponds to conventional position-resolved (camera-based) detection, whereas in the sub-diffraction regime it is naturally implemented through projections onto PSF-adapted spatial modes rather than direct pixel detection \cite{Nothlawala2025}. Thus, the present framework provides a unified description of both conventional imaging and mode-selective measurements.

\subsection{\label{sec:intro_Metro}
Metrology tools for analysis of quantum imaging schemes}

Under the assumptions introduced in the previous section, the imaging problem reduces to the estimation of the singular values $\boldsymbol{\tau}=\{\tau_k\}_k$ of the object, resulting into a multiparameter quantum estimation problem. A central concept in quantum multiparameter estimation theory is the quantum Cramér-Rao bound (QCRB). 
It states that, for a parameter-dependent quantum state $\rho_{\boldsymbol{\tau}}$, the covariance matrix $\sigma(\boldsymbol{\tau})$ of any unbiased estimator of the parameters $\{\tau_k\}_k$ satisfies the chain of matrix inequalities \cite{Liu2020}
\begin{equation}
\label{eq:QCRB}
\sigma(\boldsymbol{\tau})
\geq
\mathcal{F}^{-1}[\rho_{\boldsymbol{\tau}}, \{\hat \Omega_k\}_k]
\geq
\mathcal{Q}^{-1}[\rho_{\boldsymbol{\tau}}],
\end{equation}
where $\mathcal{F}[\rho_{\boldsymbol{\tau}}, \{\hat \Omega_k\}_k]$ is the Fisher information (FI) matrix associated with a positive operator-valued measure (POVM) $\{\hat \Omega_k\}_k$, whose elements are given by \cite{Liu2020}
\begin{equation}
\label{eq:FIM}
\mathcal{F}_{nm}[\rho_{\boldsymbol{\tau}}, \{\hat \Omega_k\}_k]=\sum_k\frac{(\partial_{\tau_n}p_k)(\partial_{\tau_m}p_k)}{p_k},
\end{equation}
with measurement probabilities determined by Born's rule as $p_k=\mathrm{Tr}[\rho_{\boldsymbol{\tau}}\hat \Omega_k]$.

The second inequality in Eq.~\eqref{eq:QCRB} expresses a measurement-independent lower bound on the achievable estimation precision, where $\mathcal Q[\rho_{\boldsymbol{\tau}}]$ denotes the quantum Fisher information (QFI) matrix. 
The first inequality can always be saturated asymptotically by choosing an efficient estimator. 
In contrast, the second inequality is not always saturable in multiparameter estimation, since a single measurement is generally not optimal for estimating all parameters simultaneously. 
This incompatibility is quantified by the compatibility matrix, $\mathcal {C}[\rho_{\bm \tau}]$, whose vanishing is a necessary and sufficient condition for asymptotic saturability \cite{Liu2020}.

For general mixed states with spectral decomposition $\rho_{\boldsymbol{\tau}}=\sum_j r_j\ket{\phi_j}\bra{\phi_j}$, the QFI matrix and compatibility matrix are given by \cite{Liu2020}

\begin{equation}
\mathcal Q_{nm}[\rho_{\bm \tau}]
=
2\sum_{\substack{j,k\\ r_j+r_k>0}}
\frac{
\mathrm{Re}
\!\left[
\bra{\phi_j}\partial_{\tau_n}\rho_{\boldsymbol{\tau}}\ket{\phi_k}
\bra{\phi_k}\partial_{\tau_m}\rho_{\boldsymbol{\tau}}\ket{\phi_j}
\right]
}{
r_j+r_k
},
\label{eq:QFI_mixed}
\end{equation}

\begin{align}
\mathcal C_{nm}[\rho_{\bm \tau}]
&=
2\sum_{\substack{j,k\\ r_j+r_k>0}}
\frac{
r_j-r_k
}{
(r_j+r_k)^2
} \nonumber \\ 
& \quad \times \mathrm{Im}
\!\left[
\bra{\phi_j}\partial_{\tau_n}\rho_{\boldsymbol{\tau}}\ket{\phi_k}
\bra{\phi_k}\partial_{\tau_m}\rho_{\boldsymbol{\tau}}\ket{\phi_j}
\right].
\label{eq:compatibility_mixed}
\end{align}

For pure states, $\rho_{\boldsymbol{\tau}}=\ket{\psi_{\boldsymbol{\tau}}}\bra{\psi_{\boldsymbol{\tau}}}$, these expressions simplify to
\begin{subequations}
\begin{align}
\mathcal{Q}_{nm}[\psi_{\bm \tau}]
&=
\mathrm{Re}
\{
\langle\partial_{\tau_n}\psi_{\boldsymbol{\tau}}
|
\partial_{\tau_m}\psi_{\boldsymbol{\tau}}\rangle \label{eq:QFI_pure}\\
&\qquad-
\langle\partial_{\tau_n}\psi_{\boldsymbol{\tau}}
|
\psi_{\boldsymbol{\tau}}\rangle
\langle
\psi_{\boldsymbol{\tau}}
|
\partial_{\tau_m}\psi_{\boldsymbol{\tau}}
\rangle
\}, \nonumber
\\
\mathcal{C}_{nm}[\psi_{\bm \tau}]
&=
\mathrm{Im}
\left\{
\langle
\partial_{\tau_n}\psi_{\boldsymbol{\tau}}
|
\partial_{\tau_m}\psi_{\boldsymbol{\tau}}
\rangle
\right\}.
\label{eq:compatibility_pure}
\end{align}
\end{subequations}

\section{Quantum Limits of Biphoton Imaging Schemes}\label{sec:All_img}

In this section, we combine the Schmidt‑mode decomposition of SPDC states, with the metrological tools and the singular‑mode object description introduced above, to the most common biphoton imaging schemes.

\subsection{\label{sec:GI}Ghost imaging}

In ghost imaging, only the signal photon interacts with the object (see Fig.\ref{fig:QI_all}\textbf{a}). 
To describe this interaction, we expand the signal Schmidt modes in the basis of the object's input singular modes 
\begin{equation}
    \ket{\psi_{\rm SPDC}} = \sum_n \sqrt{\lambda_n} \left(\sum_k \eta_{kn} \ket{f_k}_s \right)\ket{v_n}_i,
\end{equation}
with $\eta_{kn} \equiv \braket{f_k}{u_n} = \int d \vec{q} f_k(\vec{q}) u_n(\vec{q})$ denoting the overlap between both modes.
After the interaction with the object, described by Eq.~\eqref{eq:object}, the state that represents the imaging process becomes 
\begin{align}
  |\psi_\textrm{GI}\rangle =
  \sum_{n,k=0}^{\infty} \sqrt{\lambda_n}\eta_{kn}\left(\tau_k|g_k\rangle_s+\sqrt{1-\tau_k^2}\ket{e_k}\right)|v_n\rangle_i.
\end{align}

For this state, the QFI matrix resulting from Eq.~\eqref{eq:QFI_pure} takes the diagonal form
\begin{equation} \label{eq:QFI_GI}
\mathcal{Q}_{k\ell}[\psi_\textrm{GI}]= \frac{4\pi_k}{1-\tau_k^2} \delta_{k\ell},
\end{equation}
where we define $\pi_k \equiv \sum_n \lambda_n |\eta_{kn}|^ 2$ as the probability that the signal photon occupies the object singular mode $f_k$. This implies that the estimation of the transmission coefficients of the object can be independently quantified for each singular mode.
Furthermore, it follows from Eq.~\eqref{eq:compatibility_pure} that $\mathcal{C}_{k\ell}[\psi_\textrm{GI}]=0$, which means that there exists a single measurement that simultaneously achieves the QCRB for all transmission coefficients.

\subsubsection{Effective state without the environment}

In many practical situations, we do not have access to the environmental modes. Accordingly, the relevant sensitivity benchmark is not Eq.~\eqref{eq:QFI_GI}, but the QFI matrix of the reduced state $\rho_{\rm GI}^{\rm (out)} = \Tr_e [\ket{\psi_{\rm GI}}\bra{\psi_{\rm GI}}]$. 
To compute this state, we separate $\ket{\psi_{\rm GI}} \equiv \ket{\psi_T} + \ket{\psi_L}$ into its transmitted and lost parts 
\begin{subequations}
\begin{align}
\ket{\psi_T} &= \sum_{nk} \sqrt{\lambda_n}\eta_{nk}\tau_k \ket{g_k}_s \ket{v_n}_i,\\
\ket{\psi_L} &= \sum_{nk} \sqrt{\lambda_n}\eta_{nk}\sqrt{1-\tau^2_k} \ket{e_k}_s \ket{v_n}_i.
\end{align}
\end{subequations}
Because the trace over the environmental modes only involves the lost part, it leads to
\begin{equation}\label{eq:rho_GI_out}
    \rho_{\rm GI}^{\rm (out)} = P_T\rho_T + (1-P_T)\rho_L,
\end{equation}
where we have introduced the normalized states $\rho_T \equiv \ket{\psi_T}\bra{\psi_T}/P_T$ with $P_T \equiv\braket{\psi_T} = \sum_k \pi_k \tau_k^2$, and 
\begin{equation}\label{eq:rho_L}
    \rho_L = \frac{1}{1-P_T} \sum_k (1 - \tau_k^2)\ket{\chi_k}_i\bra{\chi_k} \otimes \ket{0}_s\bra{0} ,
\end{equation}
with 
\begin{equation}\label{eq:chi_k}
    \ket{\chi_k}_i \equiv \sum_n \sqrt{\lambda_n}\eta_{nk}\ket{v_n}_i.
\end{equation}
Since $\rho_T$ and $\rho_L$ live in different photon-number subspaces, the QFI matrix Eq.~\eqref{eq:QFI_mixed} reduces to
\begin{equation}\label{eq:QFI_TL}
\mathcal{Q}\left[\rho_{\rm GI}^{\rm (out)}\right] = \mathcal{F}[P_T] + P_T \mathcal{Q}[\psi_T] + (1-P_T)\mathcal{Q}[\rho_L],
\end{equation}
where we denoted with $ \mathcal{F}[P_T]$ the classical FI matrix of the transmission probabilities $\{P_T, 1-P_T\}$, given by
\begin{equation}\label{eq:QFI_GI_class}
    \mathcal{F}_{k\ell}[P_T] = 4 \pi_k \pi_\ell \tau_k \tau_\ell \left(\frac{1}{P_T} + \frac{1}{1-P_T}\right),
\end{equation}
while the quantum contribution from the pure state $\ket{\psi_T}$ is obtained from Eq.~\eqref{eq:QFI_pure}, such that
\begin{equation}\label{eq:QFI_GI_quant}
P_T \mathcal{Q}_{k\ell}[\psi_T] = 4 \pi_k \delta_{k\ell} - \frac{4 \pi_k \pi_\ell \tau_k \tau_\ell}{P_T}.
\end{equation}
We see from Eqs.~\eqref{eq:rho_L} - \eqref{eq:QFI_TL} that $\rho_L$ is also parameter dependent and accordingly it also contributes to the QFI matrix. 
This contribution can be computed using the mixed state QFI matrix Eq.~\eqref{eq:QFI_mixed}.
However, the idler only subspace is generally not used in experiments since it is very susceptible to external noise sources. 
Therefore, we will discard this contribution to the QFI matrix.
This is equivalent to considering the effective state $\rho_{\rm GI}^{\rm (eff)} = P_T \rho_T + (1-P_T)\rho_{NT}$, where $\rho_{NT}$ is a parameter-independent normalized quantum state that only keeps track of the signal losses.
Finally, combining Eqs.~\eqref{eq:QFI_GI_class} and \eqref{eq:QFI_GI_quant} we obtain 
\begin{equation}
    \mathcal{Q}_{k\ell}\left[\rho_{\rm GI}^{\rm (eff)}\right] = 4 \pi_k \delta_{k\ell} + \frac{4\pi_k \pi_\ell \tau_k \tau_\ell}{1 - P_T}.
    \label{eq:QFI_rho_eff}
\end{equation}
This result is consistent with the expectation that tracing out the environmental modes induces correlations between the parameters.
On the other hand, since the parameter dependence of the discarded subspace is retained solely through the classical transmission probability $P_T$, the effective state inherits the compatibility properties of the transmitted pure state $\ket{\psi_T}$.

Restricting to the populated subspace (where $\pi_k>0$), we apply the Sherman–Morrison formula \cite{Bartlett51} to Eq.~\eqref{eq:QFI_rho_eff} and obtain the covariance matrix  
\begin{equation}\label{eq:s_GI_eff}
\sigma_{k\ell}[\rho_{\rm GI}^{\rm (eff)}]=\Big\{\mathcal{Q}^{-1}[\rho_{\rm GI}^{\rm (eff)}]\Big\}_{k\ell}=\frac{1}{4}\frac{\delta_{k\ell}}{\pi_k} - \frac{\tau_k\tau_\ell}{4}.
\end{equation}
In the case where the Schmidt basis of the photon source matches the object basis, the probability of occupying object mode $k$ is given by the Schmidt coefficient, \textit{i.e.} $\pi_k = \lambda_k$. 

\subsubsection{Classical Fisher information for coincidence measurements}

Following the experimental implementations of ghost imaging, we now consider a coincidence measurement in which the signal is mode resolved, while the idler is detected with a bucket detector, \textit{i.e.}
\begin{equation}\label{eq:POVM_GI}
\hat \Omega_k^{\rm (c)} = \ket{g_k}_s\bra{g_k}\otimes \hat\Pi^ {(1)}_i, \quad \hat\Omega^{\rm (nc)}_k = \ket{e_k}_s\bra{e_k}\otimes \hat\Pi^ {(1)}_i.
\end{equation}
Accordingly, the probability of a coincidence in a given mode is given by $p_k^{\rm (c)} = \pi_k \tau_k^2$, and for not a coincidence in the same mode by $p_k^{\rm (nc)} = \pi_k(1-\tau_k^2)$. 
Substituting those expressions into Eq.~\eqref{eq:FIM}, the classical FI matrix is found to be 
\begin{eqnarray}
\mathcal{F}_{k\ell}[\psi_{\rm GI}, \{\hat\Omega^{\rm (c)}_k,\hat\Omega^{\rm (nc)}_k\}] = \frac{4\pi_k}{1-\tau_k^2}\delta_{k\ell} = \mathcal{Q}_{k\ell}^\textrm{GI}.
\end{eqnarray}
Therefore, this coincidence measurement achieves the ultimate quantum limit given by Eq.~\eqref{eq:QFI_GI}. However, it should be noted that this is an unusual coincidence measurement, since it records in which modes the photons have been lost. In a more common experimental scenario, we do not have access to this information. 
Therefore, a more realistic picture is described when the non coincidence POVM elements $\hat\Omega^{\rm (nc)}_k$ are replaced by a single POVM element $\hat\Omega^{\rm (nc)} = \mathds{1} - \sum_k  \hat\Omega^{\rm (c)}_k$, which leads to the non-coincidence probability
\begin{equation}
    p^{\rm (nc)} = 1- \sum_k \pi_k \tau_k^2 = 1 - P_T.
\end{equation}
This results in the non-diagonal FI matrix
\begin{equation}
    \mathcal{F}_{k\ell}[\psi_{\rm GI}, \{\hat\Omega^{\rm (c)}_k,\hat\Omega^{\rm (nc)}\}] = 4 \pi_k \delta_{k\ell} + \frac{4\pi_k \pi_\ell \tau_k \tau_\ell}{1 - P_T}, 
\end{equation}
with a diagonal contribution coming from the detection and the correlation term induced by the unresolved no-coincidence events. Noteworthy, this coincides with the QFI matrix in Eq.~\eqref{eq:QFI_rho_eff}, meaning that this coincidence measurement optimally exploits the information of the transmitted state.

\subsection{\label{sec:TP}Two-photon imaging}

In two-photon imaging, both photons interact with the object (see Fig.~\ref{fig:QI_all}\textbf{b}).
In the object-mode basis, the interaction acts diagonally and independently on both photons, according to Eq.~\eqref{eq:object}. 
Consequently, the state can be written as
\begin{align}
|\psi_\textrm{TP}\rangle
=& \sum_{k,\ell=0}^{\infty} \xi_{k\ell}\left(\tau_k|g_k\rangle_s+\sqrt{1-\tau_k^2}\ket{e_k}_s\right)\notag\\
&\qquad\times\left(\tau_\ell|g_\ell\rangle_i+\sqrt{1-\tau_\ell^2}\ket{e_\ell}_i\right),
\end{align}
where we define $\xi_{k\ell}\equiv\sum_{n=0}^\infty \sqrt{\lambda_n}\eta^{(s)}_{kn}\eta^{(i)}_{\ell n}$, with $\sum_{k,\ell=0}^\infty|\xi_{k\ell}|^2 =1,$ and $\eta_{kn}^{(s)}\equiv \braket{f_k}{u_n},  \eta_{kn}^{(i)}\equiv \braket{f_k}{v_n}$.

As we did for GI, we define the probabilities of each photon to be in mode $k$
\begin{eqnarray}
\pi_k^{(s)} \equiv \sum_{\ell} |\xi_{k \ell}|^2,\quad&
\pi_k^{(i)} \equiv \sum_{k} |\xi_{\ell k}|^2,
\end{eqnarray}
and using Eq.~\eqref{eq:QFI_pure} we obtain the QFI matrix
\begin{equation}\label{eq:QFI_TP}
\mathcal{Q}_{k\ell} [\psi_{\rm{TP}}]
=\frac{4\,\left[\pi_k^{(s)}+\pi_k^{(i)}\right]}{1-\tau_k^2}\delta_{k\ell}.
\end{equation}
Note that if the Schmidt basis matches the object basis, then
\(\xi_{k\ell}=\sqrt{\lambda_k}\delta_{k\ell}\), \(\pi_k^{(s)}=\pi_k^{(i)}=\lambda_k\), and the QFI matrix simplifies even further. 
Additionally, because all the overlaps calculated for the latter are purely real, the compatibility matrix is exactly zero.

\subsubsection{Effective state without the environment}
Let us now consider the practical case where we have no access to the environment. After tracing out all  environmental modes, the system state can be decomposed into four orthogonal blocks:
\begin{equation}
\rho_{\rm TP}
= P_{TT}\rho_{TT} + P_{TL}\rho_{TL} + P_{LT}\rho_{LT} + P_{LL}\rho_{LL},
\end{equation}
where the subscript $TT$ denotes both photons transmitted, $TL$ a transmitted signal but lost idler, similarly $LT$ a lost signal and a transmitted idler, and finally $LL$ when both are lost.

The fully transmitted block is given by $\rho_{TT} = \ket{\psi_{TT}}\bra{\psi_{TT}}$ with 
\begin{equation}
\ket{\psi_{TT}}
=\sum_{k,\ell}\xi_{k\ell}\,\tau_k\tau_\ell\,\ket{g_k}_s\ket{g_\ell}_i,
\end{equation}
with normalization factor \begin{equation}
P_{TT}
= \braket{\psi_{TT}}{\psi_{TT}}= \sum_{k,\ell}|\xi_{k\ell}|^2 \tau_k^2 \tau_\ell^2.
\end{equation}

In analogy with the GI treatment (see Eq.~\eqref{eq:rho_GI_out}), we define an effective state
\begin{equation}
\rho_{\rm TP}^{(\rm eff)} = P_{TT}\rho_{TT} + (1-P_{TT})\rho_{\rm rest},
\end{equation}
where $\rho_{\rm rest}$ is taken to be parameter-independent. Then, similarly to Eq.~\eqref{eq:QFI_TL}, the QFI matrix for the parameter vector
$\boldsymbol{\tau} = \{\tau_k\}_k$ is described by
\begin{equation}
\mathcal{Q}_{mn}\left[\rho_{\rm TP}^{(\rm eff)}\right]
= \mathcal{F}_{mn}[P_{TT}] + P_{TT}\,\mathcal{Q}_{mn}[\rho_{TT}].
\end{equation}
Combining the classical and quantum contributions, using Eqs.~\eqref{eq:FIM} and \eqref{eq:QFI_pure}, and simplifying, 
\begin{align}
\label{eq:QFI_TP_eff}
\mathcal{Q}_{mn}\left[\rho_{\rm TP}^{(\rm eff)}\right]
=&
4\,\delta_{mn}\,C_m
+ 4\,\tau_m\tau_n\big(|\xi_{mn}|^2 + |\xi_{nm}|^2\big)\notag\\
&+ \frac{4\,\tau_m\tau_n}{1-P_{TT}}\,C_mC_n,
\end{align}
with
\begin{equation}
C_k\equiv A_k+B_k, \quad A_k \equiv \sum_\ell |\xi_{k\ell}|^2\,\tau_\ell^2,\quad
B_k \equiv \sum_\ell |\xi_{\ell k}|^2\,\tau_\ell^2.
\end{equation}

When the object modes coincide with the Schmidt modes, again $\xi_{k\ell}=\sqrt{\lambda_k}\delta_{k\ell}$, which leads to $A_k=\lambda_k\tau_k^2 =B_k$ and $P_{TT}=\sum_k\lambda_k\tau_k^4$,
such that $|\xi_{mn}|^2+|\xi_{nm}|^2=2\lambda_m\delta_{mn}$. 
Therefore, for this scenario, the QFI matrix simplifies to
\begin{equation}\label{eq:QFI_TP_eff_basis_matching}
\mathcal{Q}_{mn}\left[\rho_{\rm TP}^{(\rm eff)}\right] = 16\lambda_m\tau_m^2\,\delta_{mn} + \frac{16\,\lambda_m\lambda_n\,\tau_m^3\tau_n^3}{1-\sum_k\lambda_k\tau_k^4}.
\end{equation}
In this case, restricting to the $\lambda_k>0$ subspace ({\it i.e.} the populated modes), we can use the Shermann-Morrison formula \cite{Bartlett51} to compute the covariance matrix
\begin{equation}\label{eq:s_TP_eff}
\sigma_{k\ell}\left[\rho_{\rm TP}^{\rm (eff)}\right]=\Big\{\mathcal{Q}^{-1}[\rho_{\rm TP}^{\rm (eff)}]\Big\}_{k\ell}=\frac{1}{16}\frac{1}{\lambda_k\tau_k^2}\delta_{k\ell}-\frac{\tau_k\tau_\ell}{16}.
\end{equation}

\subsubsection{Classical Fisher information for mode-resolved coincidence measurements}

As illustrated in Fig.\ref{fig:QI_all}\textbf{b}, the coincidence scheme for this imaging technique uses mode-resolved detection for both photons, in contrast to ghost-imaging, where the coincidence measurement employs a bucket detector (see Eq.~\eqref{eq:POVM_GI}). Therefore, for two-photon imaging, the coincidence POVM is $\hat\Omega_{k\ell}^{\rm (mc)} = \ket{g_k g_\ell}\bra{g_k g_\ell}$, leading to the probabilities $p_{kl}^{\rm (mc)} =  |\xi_{k\ell}|^2 \tau_k^2 \tau_\ell^2$,
and an effective non-coincidence outcome 
$p^{\rm (nc)} = 1 - \sum_{k,\ell} p_{k\ell}^{\rm (c)}= 1 - P_{TT}$. Therefore, following Eq.~\eqref{eq:FIM}, the FI matrix is computed to be
\begin{align}
\mathcal{F}_{mn}\left[ {\psi_\mathrm{TP},\{\hat\Omega_k^{\rm (mc)},\hat\Omega^{\rm (nc)} \}}\right]&
= 4\,\delta_{mn}C_m\notag\\
& +4\tau_m\tau_n\big(|\xi_{mn}|^2 + |\xi_{nm}|^2\big)\notag\\
& +\frac{4\,\tau_m\tau_n}{1-P_{TT}}\,C_mC_n.
\end{align}
Because the latter coincides with the QFI for the effective state (see Eq.~\eqref{eq:QFI_TP_eff}), it means that this measurement is quantum optimal (see Eq.~\eqref{eq:QCRB}) in the scenario where we have only access to the transmitted photons.

\subsection{\label{sec:UP}Imaging with undetected photons}

Imaging with undetected photons (UP) involves two equally probable SPDC photon-pairs generated from identical crystals (a folded setup is shown in Fig.\ref{fig:QI_all}\textbf{c}). Accordingly, the state includes a first term after one crystal (1) and object interaction via the signal-photon (see Eq.~\eqref{eq:object}), and a second crystal (2) emission probability (see Eq.~\eqref{eq:SPDC_Schmidt}), with a possible relative phase difference $\Delta\phi$. Therefore, for this setup the state is described by
\begin{align}
|\psi_\textrm{UP}\rangle =&\frac{1}{\sqrt{2}}\sum_{k=0}^{\infty} \Big(\tau_k|g_k\rangle_{s_1}\ket{\chi_k}_{i_1}\\
&\qquad +\sqrt{1-\tau_k^2}\ket{e_k}\ket{\chi_k}_{i_1}  +e^{j\Delta\phi}\ket{g_k}_{s_2}\ket{\chi_k}_{i_2}\Big), \notag
\end{align}
where we used the compacted previous definition of the idler states from Eq.~\eqref{eq:chi_k}.

Generally, these idler states are neither orthogonal nor normalized, which prevents an analytical calculation of the QFI matrix. To obtain closed-form expressions, we therefore restrict our analysis to the case where the object basis coincides with the signal Schmidt basis, \textit{i.e.} $\eta_{kn} = \delta_{kn}$, for which $\ket{\chi_k} = \sqrt{\lambda_k}\ket{v_k}$.

\subsubsection{Output state without the environment}
Similarly to the previous analyses, since the environmental modes are not accessible, we consider the state after tracing them out and aligning the signal modes of crystal 1 with those of crystal 2. This results in a mixed state 
\begin{align}\label{eq:rho_UP_out}
\rho_{\rm UP}^{(\rm out)}= \frac{1}{2}\Big[\sum_{k,\ell} &\left(
\tau_k\ket{v_k}_{i_1}+ e^{j\Delta\phi}\ket{v_k}_{i_2} \right)\notag\\
&\left(
\tau_\ell\bra{v_\ell}_{i_1}+ e^{-j\Delta\phi}\bra{v_\ell}_{i_2} \right)\notag\\
&\otimes\sqrt{\lambda_k\lambda_\ell}\ket{g_k}_{s}\bra{g_\ell}_{s} \notag\\
&\quad +\sum_k (1-\tau_k^2)\lambda_k\ket{v_k}_{i_1}\bra{v_k}_{i_1} 
\Big].
\end{align}
We can split the spectral decomposition of this quantum state into two components as 
\begin{equation}
\rho_{\rm UP}^{\rm (out)}= \zeta_0\ket{\Phi_0}\bra{\Phi_0} + \sum_k \zeta_k\ket{\Phi_k}\bra{\Phi_k}, 
\end{equation}
with eigenvalues $\zeta_0=\frac{1}{2}\sum_k \lambda_k(\tau_k^2+1)$, $\zeta_k =\frac{1}{2}\,\lambda_k(1-\tau_k^2)$,
and corresponding eigenstates
\begin{align} 
|\Phi_0\rangle &= \sum_k\frac{\sqrt{\lambda_k} \Big[ \tau_k\ket{v_k}_{i_1} + e^{j\Delta\phi}\ket{v_k}_{i_2} \Big]}{ \sqrt{2\zeta_0} }\otimes\ket{g_k}_{s}, \\
|\Phi_k\rangle &= \ket{0_{s}}\otimes\ket{v_k}_{i_1}.
\end{align}
Substituting into Eq.~\eqref{eq:QFI_mixed}, we obtain the QFI matrix
\begin{equation}
\mathcal{Q}_{nm}[\rho_{\rm UP}^{(\rm out)}]=\delta_{nm}\frac{2\lambda_n}{1-\tau_n^2}.
\label{eq:QFI_UP_SI}
\end{equation}

\subsubsection{Common only-idler measurement state}
The name imaging with undetected photons refers to the fact that only photons that did not interact with the object are detected. 
Accordingly, to faithfully represent the information accessible in experiments, we now trace out from $\rho_{\rm UP}^{(\rm out)}$ (see Eq.~\ref{eq:rho_UP_out}) all signal modes. 
The reduced state thus becomes
\begin{align}\label{eq:rho_UP_i}
\rho_{\rm UP}^{(i)}=&\frac{1}{2}\sum_k \lambda_k\Big[ \left(
\tau_k\ket{v_k}_{i_1}+ e^{j\Delta\phi}\ket{v_k}_{i_2} \right)\notag\\
&\qquad\qquad\times\left(
\tau_k\bra{v_k}_{i_1}+ e^{-j\Delta\phi}\bra{v_k}_{i_2} \right)\notag\\
&\qquad \qquad \qquad\qquad +(1-\tau_k^2)\ket{v_k}_{i_1}\bra{v_k}_{i_1} 
\Big] ,
\end{align}
which is block-diagonal in each mode $k$ with eigenvalues
$\zeta_k^\pm=\frac{\lambda_k}{2}(1\pm \tau_k)$,
and eigenstates
\begin{equation}\label{eq:UP_eigenstates_Phi}
\ket{\Phi_k^\pm}=\frac{1}{\sqrt{2}}\left(\ket{v_k}_{i_1}\pm e^{j\phi}\ket{v_k}_{i_2} \right).
\end{equation}
Consequently, using Eq.~\eqref{eq:QFI_mixed}, we obtain the QFI matrix
\begin{align}\label{eq:QFI_UP_i}
\mathcal{Q}_{mn}\left[ \rho_{\textrm{UP}}^{(i)}\right]
=\delta_{mn}\frac{\lambda_n}{1-\tau_n^2},
\end{align}
which is smaller by only a factor of two compared to Eq.~\eqref{eq:QFI_UP_SI}, where the signal information is included. 

\subsubsection{Classical Fisher information for idler measurements}

Using the reduced state $\rho_{\rm UP}^{(i)}$ (see Eq.~\eqref{eq:rho_UP_i}), which accurately describes the undetected photons setup, we now compute the classical Fisher information (see Eq.~\eqref{eq:FIM}) for measurements that erase the which-crystal information in the detection of the idler modes, \textit{i.e.} photon counting in the basis $\{\ket{\Phi^+_k},\ket{\Phi^-_k}\}_k$ (see Eq.~\eqref{eq:UP_eigenstates_Phi}).  This type of measurement was first implemented by mixing the two idler modes on a balanced beam splitter \cite{Lemos2014}, yielding two output ports that predominantly encode constructive and destructive interference, and has also been realized by aligning the paths in all degrees of freedom such that they are indistinguishable and recombined into a single output \cite{Inna2022}. Our measurement model in the $\{\Phi^\pm\}$ basis encompasses both configurations. From a metrological perspective, distributing the idlers into two outputs does not reduce the attainable Fisher information provided that both outputs are used in the data analysis, although in practice it may be more convenient to monitor a single optimized output, while a measurement with a complementary port can be reserved for applications that benefit from a separated information channel.

For single photon states, this reduces to the projective measurement $\hat\Omega^\pm_k=\ket{\Phi^\pm_k}\bra{\Phi_k^\pm}$.

The associated detection probability are then given by $p_k^\pm=\zeta_k^\pm$. 
Consequently, from Eq.~\eqref{eq:FIM} we obtain the FI matrix
\begin{align}
\mathcal{F}_{mn}\left[\rho_{\rm (UP)}^{(i)},\{\hat\Omega^\pm_k\}_k\right]
&=\delta_{mn}\frac{\lambda_n}{1-\tau_n^2},
\end{align}
which coincides with the QFI matrix in Eq.~\eqref{eq:QFI_UP_i}, revealing the optimality of this measurement.

\section{Comparison of imaging schemes}\label{sec:Comparison}

First, we note that the measurements considered throughout this work are natural mode-resolved generalizations of the detection schemes commonly employed in each imaging protocol (see Fig.~\ref{fig:SV}). We find that these measurements saturate the corresponding quantum Fisher information, implying that the standard camera-based implementations are already optimal whenever the object modes coincide with the pixel basis.

Furthermore, it is interesting to note that the QFI is proportional to the probability that photons occupy specific modes, characterized by $\pi_n$ (or simply $\lambda_n$ when the Schmidt modes match the object modes). 
Consequently, there is an intrinsic trade-off between probing many object modes and the attainable sensitivity for each of them: as more spatial modes become relevant for describing the object, the available probability is distributed among a larger number of modes, reducing the QFI associated with each transmission parameter. This is reflected in a corresponding increase in the variance of each mode parameter. Although repeated measurements can recover arbitrarily high precision for all populated modes, transmission parameters associated with unpopulated modes remain fundamentally inaccessible.

\begin{figure}[t]
\centering
\includegraphics[width=0.5\textwidth]{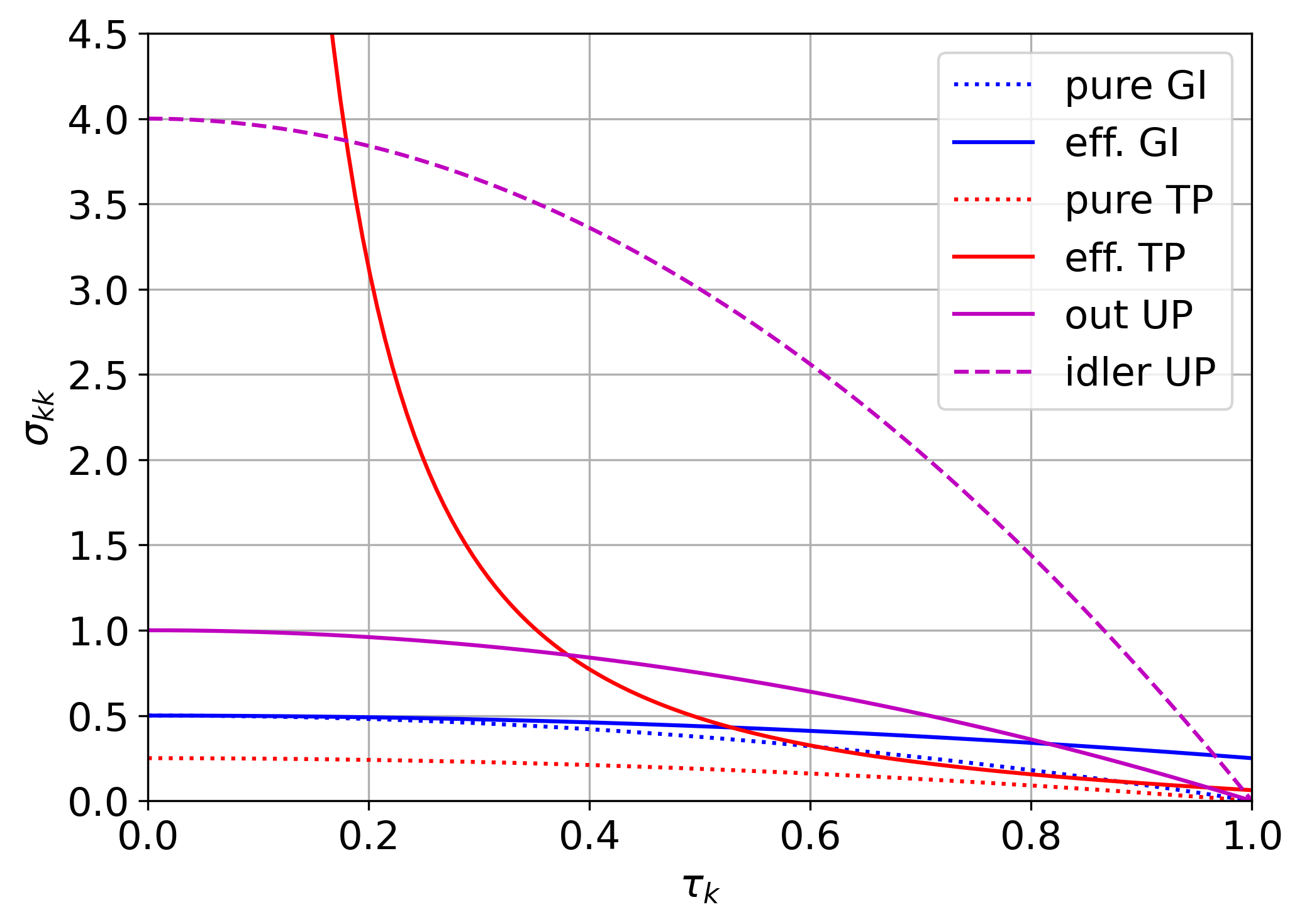}
\caption{Calculated variance $\sigma_{kk}$ (from $\{\mathcal{Q}^{-1}\}_{kk}$), for the estimation of the transmission $\tau_k$ of a single object mode using the different quantum imaging schemes. 
The blue dotted and blue solid curves show the pure-state and effective-state results for ghost imaging (GI), obtained from Eqs.~\eqref{eq:QFI_GI} and \eqref{eq:s_GI_eff}, respectively.
The red dotted and red solid curves show the corresponding results for two-photon imaging (TP), from Eqs.~\eqref{eq:QFI_TP} and \eqref{eq:s_TP_eff}. 
For imaging with undetected photons (UP), the solid purple curve corresponds to the output state after tracing out the environmental modes [Eq.~\eqref{eq:QFI_UP_SI}], while the dashed purple curve corresponds to the reduced idler state [Eq.~\eqref{eq:QFI_UP_i}].
Here we assume matching between the object and Schmidt modes ($\pi_k = \lambda_k$) and chose $\lambda_k = 0.5$ as an illustrative value.
}
\label{fig:STD_all}
\end{figure}

In Fig.~\ref{fig:STD_all}, we plot the variances $\sigma_{kk}$  obtained from the inverses of the QFI matrices as functions of the object transmission $\tau_k$ for all imaging techniques considered in the simpler scenario when the object modes coincide with the Schmidt modes ($\pi_k = \lambda_k$). 
We recall that the QFI matrix is only invertible when we restrict to the subspace of populated modes, \textit{i.e.} $\lambda_k >0$.
Moreover, $\lambda_k=1$ corresponds to the case where only one mode is populated, accordingly the eventual correlation between parameters vanishes, but solely because only a single parameter can be estimated. Accordingly, the relevant multiparameter cases are those for which $0 < \lambda_k < 1$. 
In Fig.~\ref{fig:STD_all}, we choose $\lambda_k=0.5$ as an illustrative example.
For GI we considered the inverse of the pure state QFI matrix in Eq.~\eqref{eq:QFI_GI} (dotted blue curve) and the variance for the effective state in Eq.~\eqref{eq:s_GI_eff} (solid blue curve). 
While the two coincide for small transmissions, the correlations with other transmission channels increase the estimator variance of the effective state for larger transmission values.
The latter is the scheme that shows the smallest variation over the estimation variance along the transmission range.
For TP imaging, the pure state result follows Eq.~\eqref{eq:QFI_TP} (red dotted curve), which is a factor of two below the GI variance. 
On the other hand, the variance for the effective state in Eq.~\eqref{eq:s_TP_eff} (red solid curve) remains low for high transmission, but it quickly grows for low values of $\tau_k$, for which significant information leeks to the lost-photon space.
This behavior contrasts with GI, where only one photon probes the object and the second photon remains available as a reference, making the estimation significantly more robust against low transmission.
Finally, for the UP setup we plot the QFI after tracing out the environmental modes from Eq.~\eqref{eq:QFI_UP_SI} (solid purple curve) and only retaining the idler modes from Eq.~\eqref{eq:QFI_UP_i} (dashed purple curve).
These variances are generally higher than those of all other schemes except for the effective TP scheme at low transmission.

Comparing Eqs.~\eqref{eq:s_GI_eff} and \eqref{eq:s_TP_eff}, we find that the inter-mode coupling is strongest for GI. This behavior can be directly understood from the structure of the corresponding optimal measurement: the photon probing the object is detected with a bucket detector that sums all object modes incoherently. In contrast, in TP each coincidence event is mode-resolved, so information about different object modes is more clearly separated and the inter-mode coupling is reduced. In the UP scheme, only object modes with the same index interfere. Consequently, each mode's transmission can be estimated independently and the corresponding covariance matrix is fully diagonal. These results are directly relevant to recent experimental demonstrations of quantum ghost imaging~\cite{Valerio2022} and undetected-photon imaging~\cite{Joshua2025}, which implemented the measurement schemes in a scanning configuration and do not exploit spatial correlations in the SPDC source. Our theoretical framework could be used to identify when such scanning implementations, lacking access to inter-pixel photon-pair correlations, are equivalent to camera-based schemes that do exploit these correlations.

It is important to emphasize that the optimal measurements identified above are performed in the object-mode basis. As mentioned in Sec.~\ref{sec:intro_obj}, the object-mode basis depends not only on the object itself, but also on the properties of the imaging system through its point-spread function. Consequently, only when the object is well resolved, the object modes coincide with the commonly used pixel basis, on which camera measurements are performed (see Fig.~\ref{fig:SV}). However, when this condition is not met, measurements must be adapted to the corresponding object-mode basis. For subdiffraction objects, this reduces to a basis adapted to the point-spread function, such as the Hermite--Gaussian basis proposed for estimating the separation between two incoherent sources \cite{Tsang2017}. In intermediate regimes, the optimal basis depends on both the object and the imaging system, and prior knowledge about the object's structure is generally required to identify an appropriate mode decomposition.

\section{Conclusions}

In summary, we derived the ultimate quantum limits for reconstructing the transmission profile of an object in the three imaging schemes considered—ghost imaging (GI), two-photon imaging (TP), and imaging with undetected photons (UP). We found that these limits are attained by measurements performed in the object-mode basis, which naturally generalize the standard detection schemes employed in each imaging configuration. For well-resolved objects, the object modes coincide with the conventional pixel basis, so that these optimal measurements reduce to the standard camera-based implementations. In this regime, the commonly employed measurement configurations are therefore quantum optimal, in the sense that their FI matrices saturate the corresponding QFI matrices.

Our comparative analysis shows that UP generally yields higher variances than GI and TP across most of the transmission range. Nevertheless, this comparison alone does not provide a complete assessment of the different imaging schemes. Effective GI displays the most stable performance across transmission values but also the strongest inter-mode coupling, whereas UP allows each transmission coefficient to be estimated independently. A more complete characterization would require a detailed analysis of the losses in each detection scheme, since their impact scales linearly for single-photon detection but quadratically for coincidence detection. Moreover, the interferometric configuration of the UP setup naturally suggests further investigation of its performance in phase-estimation tasks compared with the other techniques considered.
Noteworthy, our results are directly applicable to recent scanning implementations of GI and UP~\cite{Valerio2022,Joshua2025}, and could help to clarify under which conditions such schemes—without access to spatial correlations—are equivalent to camera-based, correlation-resolving implementations.

Finally, these imaging techniques are not restricted to transmission estimation: they have also been exploited for spectroscopy \cite{Chiuri2023,Placke2026} and have inspired related approaches such as plenoptic imaging \cite{Scagliola2020}. We therefore expect the quantum multiparameter estimation framework developed here to provide a systematic tool for analyzing and comparing a broader class of quantum imaging protocols, including these emerging applications. Related examples include the theoretical framework for spectroscopy with undetected photons developed in Ref.~\cite{Houde2026} and the quantum treatment of plenoptic imaging presented in Ref.~\cite{Hradil2025}.

\begin{acknowledgments}
This work was supported by the Fraunhofer Internal Programs under Grant No. Attract 40-09467.
\end{acknowledgments}

\nocite{*}

\bibliography{Refs}

\end{document}